\documentclass[10pt]{article}

\usepackage{amsmath,amssymb}
\usepackage{graphicx}
\usepackage[hang,small,bf]{caption}
\usepackage[subrefformat=parens]{subcaption}

\newcommand{\be}{\begin{equation}}
\newcommand{\ee}{\end{equation}}
\newcommand{\bea}{\begin{eqnarray}}
\newcommand{\eea}{\end{eqnarray}}
\newcommand{\ba}{\begin{array}}
\newcommand{\ea}{\end{array}}

\newcommand{\p}{\partial}

\newcommand{\la}{\langle}
\newcommand{\ra}{\rangle}

\newcommand{\cT}{{\cal{T}}}

\numberwithin{equation}{section}

\begin{document}

\allowdisplaybreaks

\title{Analysis of entropy production in finitely slow processes between nonequilibrium steady states \bigskip}

\author{
        Toshihiro Matsuo$^*$ 
       and
	Akihiko Sonoda$^\dagger$\bigskip
	\smallskip\\
	\footnotesize\it National Institute of Technology, Anan College, Tokushima 774-0017, Japan\\
	\footnotesize\tt $^*$matsuo@anan-nct.ac.jp, $^\dagger$sonoda@sb.anan-nct.ac.jp
	}

\date{\today}

\maketitle

\bigskip

\begin{abstract}
\noindent\normalsize

We investigate entropy production in finitely slow transitions between nonequilibrium steady states in Markov jump processes using the improved adiabatic approximation method proposed by Takahashi, Fujii, Hino and Hayakawa \cite{takahashi2020nonadiabatic}. This method provides systematic improvement of the adiabatic approximation on infinitely slow transitions from which we obtain nonadiabatic corrections and has a close relationship with the slow driving perturbation \cite{Mandal_2016, Cavina_2017}.
We analyze two types of excess entropy production of Sagawa and Hayakawa \cite{PhysRevE.84.051110} and Hatano and Sasa \cite{PhysRevLett.86.3463} as examples, and confirm that the leading adiabatic contribution reproduces the known results, and then obtain nonadiabatic corrections written in terms of thermodynamic metrics defined in protocol parameter spaces.
We also numerically study the resulting excess entropy production in a two-state system.

\end{abstract}


\setcounter{footnote}{0}
\section{Introduction}

Recently, thermodynamics and statistical physics on systems of nonequilibrium steady states (NESSs) have attracted much attention.
Rich thermodynamic structures in NESSs, like those in equilibrium thermodynamics have been revealed through studies on stochastic thermodynamics of Markov jump processes and Langevin systems, etc \cite{Seifert_2012}. 
In such studies, the concept of the housekeeping steady heat $Q_{hk}$ which is the heat transfer necessary to maintain a system in a given steady state plays the key role and the renormalized excess heat is defined as $Q_{ex}=Q_{tot}-Q_{hk}$ where $Q_{tot}$ is the total heat transfer of the system \cite{{Landauer_1978}, {10.1143/PTPS.130.29}, {PhysRevLett.86.3463}, {Ruelle_2003}, {KNST1}, {Maes_2014}}.\footnote{Recently, Dechant, Sasa and Ito have introduced in \cite{dechant2022} the third component of the decomposition named as ``the coupling entropy production" in their geometric approach which plays a role of interactions between different types of driving forces.}
Accordingly, the extended form of the Clausius relation, which is one of the cores of equilibrium thermodynamics, has been examined in order to construct a NESS version of thermodynamics \cite{Crooks_1999,Speck_2005,Seifert_2005,Sasa_2006,KNST2,Saito_2011,Pradhan_2011,PhysRevE.80.021137,PhysRevE.81.051133,Esposito_2010,RevModPhys.81.1665}.
Among these, Komatsu, Nakagawa, Sasa and Tasaki (KNST) found KNST's extended Clausius inequality using a symmetrized version of a Shannon entropy in the lowest order of nonequilibrium \cite{KNST1,KNST2}. 
Furthermore, in \cite{PhysRevE.84.051110} the geometric expressions of the excess entropy production for quasi-static transitions between NESSs have been derived. 
The expressions imply vector potentials, instead of scalar potentials, in parameter spaces play roles in steady state thermodynamics (SST). 
Nevertheless much research has concerned mainly {\it quasi-static} transitions between NESSs, finitely slow transitions beyond the regime of the adiabatic/quasi-static transitions are still elusive. 
It is necessary to investigate the aforementioned findings even on finitely slow transitions with nonadiabatic corrections in order to put forward our understandings on SST.
In \cite{takahashi2020nonadiabatic} Takahashi, Fujii, Hino and Hayakawa have introduced a method that systematically improves the adiabatic approximation of the solution of the master equation and enables us to investigate beyond the adiabatic regime.
In this paper we investigate excess entropy production of Sagawa-Hayakawa in finitely slow transition between the NESSs with the improved adiabatic approximation method. 
We also examine the Hatano-Sasa entropy production as another example.

This paper is organized as follows.
In Section 2, we introduce the Markov jump process of finite states and entropy production. 
In Section 3, we briefly review the improved adiabatic approximation method for the stochastic master equation 
and point out the relationship to the slow driving perturbation \cite{Mandal_2016, Cavina_2017}.   
We apply the method to the entropy production to get perturbative series expansions.
In Section 4, we consider a typical two-state system as a concrete example where we examine the excess entropy production with the nonadiabatic effects in slow processes.
Section 5 is devoted to discussion and summary. 

\section{Master equation of Markov jump process and entropy production}

We consider a continuous-time Markov jump process with discrete finite $N (< \infty)$ states.
Let $p_i(t)$ be a probability of the system in a state $i (=0,1, \ldots , N-1)$ at time $t$, and the probability distribution 
is given by $|p(t)\ra := (p_0(t), p_1(t), \cdots, p_{N-1}(t))^T$.
The time evolution of $|p(t) \ra$ is described by the following stochastic master equation:
\begin{align}
\frac{d}{d t}|p(t) \ra = R(\alpha(t)) |p(t) \ra ,
\end{align}
where $R(\alpha(t))$ is a $N \times N$ transition rate matrix which is a function of time-dependent parameters 
$\{\alpha_\mu(t)\}_{\mu=1,2, \cdots}$ collectively denoted as $\alpha(t)$ which describe protocols through which external agents operate the system.
The system is supposed to be in contact with multiple different reservoirs and the rate matrix composes as $R(\alpha(t))=\sum_a R^a$ with $a$ being the label of reservoirs.
The normalization of the probability distribution is written with the use of $\la 1| :=(1,1, \ldots,1)$ as
\begin{align}
\la 1 | p(t) \ra = \sum_i p_i(t) = 1 .
\end{align}
We see from the normalization of the probability distribution at arbitrary time that the transition rate matrix satisfies 
the normalization condition:
\begin{align}
(\la 1 | R)_j =\sum_i R_{ij}=0 ,
\label{xRxy}
\end{align}
which implies the existence of zero eigenvalue.
We suppose the transition rate matrix $R$ is irreducible and there is no degeneracy. 
The transition rate matrix is not hermitian thus the left and right eigenvectors do not conjugate with each other.
The solution of the master equation is expressed formally as
\begin{align}
|p(\tau)\ra 
={\cT} \exp\left(
\int^\tau_0 dt R(\alpha(t))
\right) |p(0)\ra ,
\end{align}
where $\cT$ is the time ordering.

Assuming the local detailed balance condition we define the entropy production in the reservoirs associated with the transition $j \to i$ by
\begin{align}
\sigma_{ij}^a := \ln \frac{R_{ij}^a}{R_{ji}^a} = - \beta_a Q_{ij}^a 
\label{rEP}
\end{align}
for the case $R_{ij}^a \neq0$ and $R_{ji}^a \neq0$, where $Q_{ij}^a$ is the heat absorbed from the heat reservoir with the inverse temperature $\beta_a$.
A total process is composed of a sequence of transitions $j \to i \to k \to \cdots$ forming a trajectory in the state space and the accumulated reservoir entropy production in the total process is given by
\begin{align}
\sigma_r := \sum_a \sum_t {\sigma}_{i(t+0)j(t-0)}^a ,
\end{align}
which depends on the trajectory and hence is the stochastic quantity.
The expectation value of the entropy production is given by 
\begin{align}
\la \sigma_r \ra &= \int_0^\tau dt  \dot{\sigma}_r ,
\label{resEP}
\end{align}
where the entropy production rate is given by
\begin{align}
\dot{\sigma}_r &:= \sum_a \sum_{i,j}R_{ij}^a p_j  {\sigma}_{ij}^a 
\nonumber \\
&= \la 1| R\sigma_r | p(t) \ra ,
\label{resEPrate}
\end{align}
where the ${ij}$ component of $R\sigma_r$ is given as $(R\sigma_r)_{ij}:=\sum_a R^a_{ij}\sigma^a_{ij}$.

The system entropy production rate is provided by the Shannon entropy $H = -\sum_i p_i \ln p_i$ as
\begin{align}
\dot{H}=-\sum_a \sum_{i,j}R^a_{ij}p_j \ln{p_{i} \over p_{j}} 
\end{align}
where we have used the master equation for $p_i$ and the normalization property of the rate matrix \eqref{xRxy} \cite{CoverThomas}.
The sum of the above two entropy production rates gives a total entropy production rate \cite{Seifert_2012}
\begin{align}
\dot{\sigma}_{tot} :&= \dot{\sigma}_r +\dot{H}
\nonumber \\
&=\sum_a \sum_{i,j}R_{ij}^a p_j \ln\frac{R_{ij}^a p_{j}}{R_{ji}^a p_{i}} 
\end{align}
which can be shown to be positive \cite{RevModPhys.81.1665}.

In this paper we focus on entropy production in transitions between steady states.
We consider instantaneous steady state probabilities $p_i^{S}(\alpha(t))$ with a given protocol parameter $\alpha(t)$ which is defined through the relation:
\begin{align}
R(\alpha(t)) |p^{S}(\alpha(t)) \ra =0.
\label{RpS=0}
\end{align}
Note that the instantaneous steady state $|p^S(\alpha(t)) \ra$ itself is not in general a solution of the master equation except for constant protocol $\alpha(t) =\alpha$ where the time derivative of $|p^S(\alpha) \ra$ vanishes.

Steady states require certain amounts of heat to maintain the states called housekeeping heat \cite{10.1143/PTPS.130.29} which is accompanied by housekeeping entropy production.
We consider two types of housekeeping entropy production rates.
One is defined by replacing $|p(t)\ra$ with an instantaneous steady state
$|p^S(\alpha(t))\ra$ in the reservoir entropy production rate \eqref{resEPrate}:
\begin{align}
\dot{\sigma}_{hk}&:=\sum_a\sum_{i,j}R^a_{ij}p^S_j \ln\frac{R^a_{ij}}{R^a_{ji}} 
\nonumber \\
&= \la 1| R\sigma_r | p^S(\alpha(t)) \ra .
\label{SHhk}
\end{align}
This entropy production rate has been considered in \cite{KNST1, KNST2, PhysRevE.84.051110} and we call this the Sagawa-Hayakawa housekeeping entropy production rate\footnote{In \cite{PhysRevE.84.051110}, this expression is derived as a ``dynamical pase" in a formulation using the full-counting statistics. }.
The reservoir entropy production rate is renormalized by the housekeeping entropy production rate to give the  corresponding (Sagawa-Hayakawa) excess entropy production rate:
\begin{align}
\dot{\sigma}_{ex} :&= \dot{\sigma}_{r} -\dot{\sigma}_{hk}
\nonumber \\
&= \sum_a\sum_{i,j}R^a_{ij} \ln\frac{R^a_{ij}}{R^a_{ji}} (p_j-p^S_j).
\end{align}
The excess entropy production rate $\dot{\sigma}_{ex} $ vanishes for a steady state where $|p(t) \ra=| p^S \ra$ as it should.
Note, however, that the integrated excess entropy production does not vanish in transitions between different steady states because the deviation of the intermediate state from the steady state contributes to the integral.
This is even the case in infinitely slow (or adiabatic) processes in which intermediate states can be regarded as {\it almost} steady states.

We also focus on another housekeeping entropy production rate, namely, the Hatano-Sasa housekeeping entropy production rate  \cite{PhysRevLett.86.3463}:
\begin{align}
\dot{\Sigma}_{hk} := \sum_a\sum_{i,j}R^a_{ij}p_j \ln\frac{R^a_{ij}p^S_{j}}{R^a_{ji}p^S_{i}} ,
\end{align}
which is positive definite that can be shown by using the property \eqref{RpS=0} \cite{RevModPhys.81.1665}.
The corresponding (Hatano-Sasa) excess entropy production rate is given by 
\begin{align}
\dot{\Sigma} :&= \dot{\sigma}_{r} -\dot{\Sigma}_{hk} 
\nonumber \\
&= \sum_a\sum_{i,j}R^a_{ij}p_j \ln\frac{p^S_{i}}{p^S_{j}} ,
\label{HSEPr}
\end{align}
which also vanishes for a steady state $| p(t) \ra=| p^S \ra$.
The entropy production rate also provides nonzero entropy production under integrations even in infinitely slow processes.
Indeed, it is a well known fact that the Hatano-Sasa entropy production becomes the minus Shannon entropy difference in adiabatic slow processes as a result of the Hatano-Sasa relation \cite{PhysRevLett.86.3463}.


\section{Improved adiabatic approximation}

The solutions of the master equation $| p(t) \ra$ are well described by the adiabatic approximation for slow processes in which states of the system stay steady at arbitrary time. 
Here we are interested in deviations from slow processes and how nonadiabatic effects would develop. 
In \cite{takahashi2020nonadiabatic} Takahashi, Fujii, Hino and Hayakawa have introduced a method that systematically improves the adiabatic approximation of the solution of the master equation and enables us to investigate beyond the adiabatic regime
and further study has been done in \cite{Takahashi_2020}.
In this section, we review the improved approximation method and will apply it to the formula of the entropy production in the next section.

Let $\lambda^n(\alpha)$ be the $n$-th eigenvalue of $R(\alpha)$, where $\mbox{Re} \lambda^n > \mbox{Re} \lambda^m$ for $m>n$ 
from Perron-Frobenius theorem and we set $n=0$ which corresponds to the zero eigenvalue, $\lambda^0=0$.
We assume there is no degeneracy.
The corresponding left and right eigenvectors satisfy
\begin{align}
\la \lambda^n |R = \la \lambda^n |\lambda^n, 
\quad
R|\lambda^n\ra = \lambda^n|\lambda^n\ra , 
\end{align}
and we set the normalization
\begin{align}
\la \lambda^n(\alpha)|\lambda^m(\alpha)\ra = \delta_{nm}.
\end{align}
Since the rate matrix $R(\alpha)$ is not hermitian, the left and right eigenvectors do not conjugate with each other.
We note that the states depend on time through the time-dependent protocols $\alpha(t)$ thus 
we shall abbreviate $|\lambda^n(\alpha(t))\ra$ as $|\lambda^n(t)\ra$ in the following argument.

It should be noted that there is a symmetry of arbitrariness in the set of the eigenvectors $\{ \la\lambda^n |\}, \{|\lambda^n\ra\}$
\begin{align}
\la\lambda^n | \to \la\lambda^n|S_n^{-1}(t), \quad 
|\lambda^n \ra \to S_n(t)|\lambda^n\ra ,
\label{transS}
\end{align}
where $S(t)$ is an arbitrary function.
We fix this arbitrariness \cite{takahashi2020nonadiabatic} by taking
\begin{align}
\la \tilde{\lambda}^n | :=  \la\lambda^n | e^{\gamma_n(t)}, \quad
|\tilde{\lambda}^n \ra := e^{-\gamma_n(t)} |\lambda^n\ra ,
\end{align}
where
\begin{align}
\gamma_n(t)&:=\int_0^t ds \la \lambda^n(s)|\dot{\lambda}^n(s)\ra 
\label{geometricphase}
\end{align}
is the geometric phase \cite{Berry1984, PhysRevLett.58.1593, PhysRevLett.60.2339, Sinitsyn_2007, PhysRevLett.99.220408}.
The set of states $\{\la \tilde{\lambda}^n |\}, \{|\tilde{\lambda}^n \ra\}$ have a special property, namely these are invariant under the transformation \eqref{transS}, and as a consequence
\begin{align}
\la \tilde{\lambda}^n(t) |\dot{\tilde{\lambda}}^n(t)\ra&=0, \quad(\forall n)
\end{align}
which implies the geometric phase is zero in this basis and we call this the zero phase condition.
The orthonormal condition is unchanged in this basis,
\begin{align}
\la \tilde{\lambda}^m(t) |\tilde{\lambda}^n(t)\ra &=\delta_{mn}.
\end{align}
In the following arguments, it should be understood that the states with tilde satisfy the zero phase condition.
Still there is arbitrariness about constant multiplicative factor though it does not affect the final results.

We expand the solution of the master equation $|p(t) \ra$ with respect to the basis $|\tilde{\lambda}^n(t)\ra$ as
\begin{align}
|p(t) \ra = \sum_n c_n(t) e^{\Lambda^n(t)} |\tilde{\lambda}^n(t)\ra,
\label{p(t)}
\end{align}
where 
\begin{align}
\Lambda^n(t):=\int_0^t ds \lambda^n(s) 
\end{align}
is the dynamical phase.

We obtain the differential equation for the coefficients $c_n(t)$ by substituting the above expression to the master equation:
\begin{align}
\dot{c}_n(t)
=-\sum_{m=0} c_m(t) e^{-(\Lambda^n(t)-\Lambda^m(t))}
\la \tilde{\lambda}^n(t)|\dot{\tilde{\lambda}}^m(t)\ra .
\label{dot{c}_n(t)}
\end{align}
In the ordinary adiabatic approximation in quantum mechanics, the transition rates between different level of eigenstates might be neglected. 
It follows that the terms other than $m=n$ would be omitted in equations corresponding to \eqref{dot{c}_n(t)}, and only a single level of state plays the role.
However this is not the case here because of the Euclidean nature of the set up. 
We shall employ a different approximation scheme where we keep only the term with $m=0$ and omit the terms with positive $m$ in the sum which is justified by the presence of the exponential suppressions.
It follows that after the time integration we find
\begin{align}
c_n(\tau)
&\simeq
c_n(0)-\int_0^\tau dt 
e^{-\Lambda^n(t)}
\la \tilde{\lambda}^n(t)|\dot{\tilde{\lambda}}^0(t)\ra .
\label{cntau}
\end{align}

Note that the zero eigenvalue states $\la 1|$ and $| p^S \ra$ satisfy the zero phase condition $\la 1 | \dot p^S\ra =0$ which is derived by differentiating the normalization condition $\la 1 | p^S \ra =1$.
Then we identify $\la \tilde{\lambda}^0(\alpha) |=\la 1|$, $|\tilde{\lambda}^0(\alpha)\ra =|p^S(\alpha)\ra$
and we have $c_0(t)=\la\tilde{\lambda}^0|p(t)\ra=\la 1|p(t)\ra=1$ for $\forall t$.
Furthermore, we consider processes in which systems start from steady state distributions and develop according to given protocols $\alpha(t)$. 
This implies the initial condition $c_n(0)=0$ for $n\neq0$.
Putting \eqref{cntau} back into \eqref{p(t)} we obtain the approximated solution of the master equation in the following form:
\begin{align}
|p(\tau) \ra &\simeq
|p^S(\alpha(\tau))\ra+ \sum_{n=1}
|\tilde{\lambda}^n(\tau)\ra  \delta_n(\tau),
\label{ptau}
\end{align}
where 
\begin{align}
\delta_n(\tau) = -\int_0^\tau dt 
e^{\int_t^\tau dt' \lambda^n(t')} 
\la \tilde{\lambda}^n(t)|\dot{\tilde{\lambda}}^0(t)\ra .
\label{delta}
\end{align}
The second term in \eqref{ptau} contains the factor which is omitted in the ordinary adiabatic approximation as a nonadiabatic mixing effect.
This term may be further decomposed in the following way.
It is easy to see the coefficient $\delta_n(\tau)$ satisfies the following equation:
\begin{align}
\frac{d \delta_n(t)}{d t}
&=
\lambda^n(t)
\left[
\delta_n(t) 
-\frac{
\la \tilde{\lambda}^n(t)|\dot{\tilde{\lambda}}^0(t)\ra
}{ 
\lambda^n(t)}
\right] .
\end{align}
Since the eigenvalues $\lambda^n(t) (n\neq0)$ are negative, the $\delta_n(t)$ quickly approaches the fixed point:
\bea
\delta_n^{(0)}(t) 
:={
\la \tilde{\lambda}^n(t)|\dot{\tilde{\lambda}}^0(t)\ra
\over 
\lambda^n(t)} .
\label{deltazero}
\eea
We can take iteratively the derivative expansion of $\delta_n(t)$ around the fixed point to get
\bea
\delta_n(t) 
= \sum_{k=0}^\infty \left(\frac{1}{\lambda^n(t)}\frac{d}{dt} \right)^k \delta_n^{(0)}(t) .
\eea
With this decomposition we have the improved adiabatic expansion:
\begin{align}
|p(t) \ra &\simeq
|p^S(\alpha(t))\ra
+
\sum_{n=1} \sum_{k=0}^\infty |\tilde{\lambda}^n(t)\ra
\left({1 \over \lambda^n(t)}\frac{d}{dt} \right)^k \delta_n^{(0)}(t)
.
\label{IAE}
\end{align}
This is an approximation in the sense that we neglect the transitions between different states other than the transitions to a steady state. 
Note that the diagonal transitions do not exist due to the zero phase condition in this basis.

A comment is in order here.
In \cite{Mandal_2016, Cavina_2017} slow transitions between nonequilibrium steady states have been analyzed by the slow driving perturbation method for probability distribution.
One might wonder about the relationship to the improved adiabatic approximation.
In fact, the improved adiabatic expansion is derived from the slow driving perturbation as we shall briefly explain  below. 
Let us write the probability distribution as the sum of the steady state with a small correction as
\begin{align}
|p(t) \ra &= |p^S(\alpha(t))\ra + |\Delta p(t) \ra .
\end{align}
Substituting this into the master equation we obtain 
\begin{align}
 \frac{d}{dt} |\Delta p(t) \ra &= R |\Delta p(t) \ra -\frac{d}{dt} |p^S(\alpha(t))\ra .
\label{solvedeq} 
\end{align}
To solve this equation, we introduce the Moore-Penrose generalized inverse matrix
\begin{align}
R^+:&=\sum_{n \neq 0} \lambda^{-1}_n(t) | \tilde{\lambda}^n \ra \la \tilde{\lambda}^n |
\nonumber \\
&=\int_0^\infty dt e^{R t} \left(|p^S\ra\la 1|-1\right) 
\end{align}
which satisfies 
\begin{align}
&R^+ R = R R^+ = 1 - |p^S\ra\la 1| , 
\nonumber \\
&R^+ |p^S\ra =0, \quad \la 1|R^+=0 .
\label{R^+relations}
\end{align}
Multiplying the generalized inverse matrices to \eqref{solvedeq} from the left we obtain
\begin{align}
\left(1- R^+\frac{d}{dt}\right) |\Delta p(t) \ra=R^+\frac{d}{dt} |p^S(\alpha(t))\ra .
\end{align}
Here we have used $\la 1|\Delta p(t) \ra=0 $ which is followed by the conservation of probability $\la 1|p(t) \ra=\la 1|p^S(t) \ra=1$.
This is solved iteratively to get the slow driving expansion:
\begin{align}
|p(t) \ra &= \sum_{k=0}^\infty \left(R^+ \frac{d}{dt}\right)^k |p^S(\alpha(t))\ra . 
\label{slow}
\end{align}
So far the expression is exact.\footnote{ It has been pointed out in \cite{Nakajima_2021} that the series is an asymptotic expansion though is tractable with the Borel summation.} 
We now show that this expansion is reduced to the improved adiabatic expansion \eqref{IAE}. 
First let us note the relation
\begin{align}
\la \tilde{\lambda}^n(t) | R^+ \frac{d}{dt} | \tilde{\lambda}^m(t) \ra 
=
\delta_{nm}{1 \over \lambda^n(t)}\frac{d}{dt}
+
\frac{1}{\lambda^n(t)}
\la \tilde{\lambda}^n(t) | \dot{\tilde{\lambda}}^m(t) \ra 
\quad (n \neq 0, m \neq 0) .
\label{offdiag}
\end{align}
The off-diagonal elements with $n\neq m (n,m \neq 0)$ in the second term might be neglected under the adiabatic approximation as we have done in the improved approximation.
Furthermore, the diagonal terms would vanish in the zero phase condition $\la \tilde{\lambda}^n(t) | \dot{\tilde{\lambda}}^n(t) \ra=0$.
Then we obtain the approximate relation:
\begin{align}
\la \tilde{\lambda}^n(t) | R^+ \frac{d}{dt} | \tilde{\lambda}^m(t) \ra 
\simeq
\delta_{nm}{1 \over \lambda^n(t)}\frac{d}{dt}
\quad (n \neq 0, m \neq 0) .
\end{align}
Then, by inserting the completeness relation $\sum_{n=0} |\tilde{\lambda}^n \ra \la \tilde{\lambda}^n| =1$ between the products of the derivatives in \eqref{slow}, and noting the fact that the zero modes vanish due to the condition $\la 1|R^+=0$, we finally obtain the improved adiabatic expansion \eqref{IAE}.\footnote{
It would seems that the slow driving expansions would reduce to the improved adiabatic expansion only with the particular gauge fixing condition, however this is not the case.
Even without the condition, we find the diagonal terms play no role in the calculation as should be expected by the symmetry.}
We note that the two methods are equivalent when the number of the states is less than two in which the off-diagonal terms in \eqref{offdiag} are absent.\\

Now we plug the expansion \eqref{IAE}, or \eqref{slow} into \eqref{resEPrate} to get approximated entropy production.
We obtain the Sagawa-Hayakawa housekeeping entropy production from the first term of the expansion which represents the steady state.  
We also get the Sagawa-Hayakawa excess entropy production in a form 
\begin{align}
\la \sigma \ra_{ex} = \sum_{k=0} \la \sigma \ra^{(k)}_{ex} 
\end{align}
where
\begin{align}
\la \sigma \ra^{(k)}_{ex}
&= \int_0^\tau dt \la 1 | R \sigma    \left(R^+ \frac{d}{dt} \right)^{k+1}  |p^S \ra
\nonumber \\
&\simeq \int_0^\tau dt \sum_{n=1}  \la 1 | R \sigma |\tilde{\lambda}^n(t) \ra
\left({1 \over \lambda^n(t)}\frac{d}{dt} \right)^{k} \delta_n^{(0)}(t) 
\end{align}
where we have used the slow driving expansion \eqref{slow} in the first line, and the improved adiabatic expansion \eqref{IAE} in the second line.

It is obvious that the leading ($k=0$) term is written as
\begin{align}
\la \sigma \ra^{(0)}_{ex}
&= \int \la 1 | R \sigma R^+  d  |p^S \ra 
\label{BSNphase}
\end{align}
which is geometric in the sense that it depends on the geometry of the path traversed in parameter space spanned by $\alpha^\mu$.
The geometric expressions in the adiabatic limit have been argued in various contexts \cite{
{Sinitsyn_2007}, {PhysRevLett.99.220408}, {PhysRevE.84.051110}, {articleOhkuboEggel}}.
In particular, in \cite{PhysRevE.84.051110}, the expression \eqref{BSNphase} has been obtained as the Berry-Sinitsyn-Nemenman phase in a formulation with the full-counting statistics.

As for the Hatano-Sasa entropy production \eqref{HSEPr}, 
the steady state term $|p^S \ra$ in \eqref{IAE} and \eqref{slow} does not contribute to the excess entropy production 
because of the properties of steady state probability.
Then the entropy production becomes
\begin{align}
\langle \Sigma\rangle=\sum_{k=0} \langle \Sigma\rangle^{(k)} ,
\label{HSEP}
\end{align}
where
\begin{align}
\langle \Sigma\rangle^{(k)}
&=
\int_0^\tau dt  \sum_{i,j}R_{ij} \ln p^S_i 
\left[
\left(R^+ \frac{d}{dt}\right)^{k+1} p^S(t) \right]_j
\nonumber \\
&\simeq
\int_0^\tau dt  \sum_{i,j}R_{ij} \ln p^S_i 
\sum_{n=1} \tilde{\lambda}^n_j 
\left({1\over \lambda^n(t)} \frac{d}{dt}\right)^k 
\delta^{(0)}_n(t) 
\end{align}
where again we have used the expansions \eqref{slow} and \eqref{IAE} in the first and the second lines, respectively.

It is easy to see by using the relation \eqref{R^+relations} that the leading ($k=0$) term in \eqref{HSEP} is equal to the minus of the Shannon entropy difference $\Delta H$ in quasi-static processes:
\begin{align}
\langle \Sigma\rangle^{(0)} = -\Delta H
\end{align}
where $\Delta H$ is the difference between the initial and final distributions, both of which are supposed to be steady states:
\begin{align}
\Delta H 
&:= -\sum_i p^S_i(\tau) \ln p^S_i(\tau)+\sum_i p^S_i(0) \ln p^S_i(0).
\end{align}
Since the leading term represents an adiabatic (infinitely slow) transition, we see this is a consequence of the well known Hatano-Sasa inequality.

It is interesting to see the subleading $k=1$ terms in which we may find the geometric structure for the $k=1$ term in the Hatano-Sasa entropy production \cite{Mandal_2016, PhysRevResearch.3.013187}. 
We are able to define a metric in parameter spaces as follows. 
We write the $k=1$ term in a form:
\begin{align}
\langle \Sigma\rangle^{(1)} 
&= \sum_{i,j} \ln p^S_i R^+_{ij} {d p^S_j \over dt}\Big|^\tau_0
-\int_0^\tau dt \sum_{i,j} p^S_j {d \ln p^S_i \over dt} R^+_{ij} {d \ln p^S_j \over dt} .
\end{align}
We may omit the surface term since we consider the transitions between steady states in which time derivatives vanish.
Therefore we have, recalling the protocol parameter is given by $\alpha^\mu$, 
\begin{align}
\langle \Sigma\rangle^{(1)}
&= \int_0^\tau dt g_{\mu\nu}^{HS}(\alpha) \dot{\alpha}^\mu \dot{\alpha}^\nu 
\end{align}
where 
\begin{align}
g_{\mu\nu}^{HS}(\alpha)=-{1 \over 2} \sum_{i,j} p^S_j  
\left(
{\p \ln p^S_i \over \p \alpha^\mu} R^+_{ij}  {\p \ln p^S_j \over \p \alpha^\nu}
+{\p \ln p^S_i \over \p \alpha^\nu} R^+_{ij}  {\p \ln p^S_j \over \p \alpha^\mu}
\right)
\end{align}
which is so-called the thermodynamic metric.

\section{Two-state system} 
Now let us take, as a concrete example, the two-state system which is a typical model of a quantum dot in the strong Coulomb blockade regime \cite{PhysRevB.67.085316}.
As we mentioned before, the improved adiabatic approximation in two-state systems is exactly same as the slow driving perturbation.
Therefore the word adiabatic is equivalent to quasi-static in this case.

The system is supposed to take two states $i=0$ and  $i=1$, which respectively describe that the electron is absent and occupies the dot.
The system couples to $M$ reservoirs specified by inverse temperature $\beta_a$ and chemical potential $\mu_a$ where  $a$  is the label of the reservoirs.
The transition rate is given by $R=\sum_{a=1}^M R^a$ with
\begin{align}
R^a=\left(\begin{array}{cc} - f_a & 1- f_a \\ f_a & f_a -1\end{array}\right),
\end{align}
where $f_a$ is the fermi distribution function $f_a:=( e^{\beta_a(E-\mu_a)}+1)^{-1}$
with $E$ being the excitation energy of the dot.\footnote{We have set the rates of the transition from the dot to reservoirs to one for simplicity.}  
The reservoir entropy production in the transition from the state $i=1$ to $i=0$ is
$\ln({R_{01}^a/R_{10}^a})=\sigma_{01}^a=-\sigma_{10}^a=-\sigma_a$
with $\sigma_a=-\beta_a(E-\mu_a)$, and thus 
the fermi distribution function is written as $f_a=(e^{-\sigma_a}+1)^{-1}$. 
The $ij$ component of the product of the transition rate and the reservoir entropy production is
\begin{align}
(R^a \sigma^a)_{ij} &=\left(\begin{array}{cc} 0 & -(1- f_a)\sigma_a \\ f_a \sigma_a  & 0 \end{array}\right) .
\end{align}
The two eigenvalues of the transition rate $R$ are $\lambda^0=0$ and $\lambda^1=-M$ and corresponding right eigenvectors are 
\begin{align}
|\tilde{\lambda}^0 \ra =|p^S \ra= {1\over M} \left(\begin{array}{c}  M - F  \\  F \end{array}\right) ,
\qquad 
|\tilde{\lambda}^1 \ra= \left(\begin{array}{c}  1 \\  -1 \end{array}\right)
\end{align}
respectively, where $F=\sum_a f_a$.
For left eigenvectors we have 
\begin{align}
\la \tilde{\lambda}^0 | = \la 1 |=(1,1), \quad 
\la \tilde{\lambda}^1 |={1\over M} (F, F-M) .
\end{align}
One may check that the geometric phase in each state is zero, hence we have put the tilde on the states.
We find the fixed point function \eqref{deltazero} appearing in the expansion of the improved adiabatic approximation:
\begin{align}
\delta^{(0)}_1(t) &=
{1 \over M^2} \dot{F} .
\end{align}

Therefore the distribution function \eqref{IAE} takes a form as
\begin{align}
|p(t) \ra = {1\over M} \left(\begin{array}{c}  M -  F \\  F \end{array}\right)
+ \sum_{k=0}^\infty\left(-{1 \over M}\right)^{k+2}{d^{k+1} F\over dt^{k+1}} 
\left(\begin{array}{c}  1 \\ -1 \end{array}\right) .
\label{pket}
\end{align}
Plugging (\ref{pket}) into the formulae in the previous section, we obtain the entropy production. 

The Sagawa-Hayakawa housekeeping entropy production \eqref{SHhk} is obtained from the steady state term as
\begin{align}
\dot{\sigma}_{hk}(t)
&=
\sum_a f_a \sigma_a -{1\over M} \sum_a f_a \sum_b\sigma_b ,
\end{align}
which provides the housekeeping entropy production:
\begin{align}
\la \sigma \ra_{hk}
&= \int_0^\tau dt \dot{\sigma}_{hk}(t) .
\label{EPad}
\end{align}
Then we get the excess part of the entropy production:
\begin{align}
\la \sigma \ra_{ex}^{(k)} 
&=
\left(-{1\over M}\right)^{k+2}
\int^\tau_0 dt
\sum_a \sigma_a {d^{k+1} F \over dt^{k+1}} .
\label{EPnak}
\end{align}
In particular, the leading $k=0$ term provides, by noting $\dot{F} =  \sum_a f_a(1-f_a)\dot{\sigma}_a$, 
\begin{align}
\la \sigma \ra_{ex}^{(0)} 
&={1\over M^2}\int_C \sum_{a} \sigma_a
\sum_b f_b(1-f_b)d\sigma_b .
\label{EPgeo}
\end{align}
This result has been derived originally in \cite{PhysRevE.84.051110} in the method of the full counting statistics where the derivation is quite similar way as we get the geometric phase in quantum mechanics \cite{Berry1984,PhysRevLett.58.1593}.\footnote{The integrand in (\ref{EPgeo}) itself does not transform as a vector field. The gauge field lives in the space which is different from that spanned by $\sigma_a$.}

We write the $k=1$ term in the Sagawa-Hayakawa entropy production as 
\begin{align}
\la \sigma \ra_{ex}^{(1)}
&=
-{1 \over M^3} \int_0^\tau dt \sum_a \sigma_a {d^{2} F\over dt^{2}} 
\nonumber \\
&=
-{1 \over M^3} \sum_a \sigma_a {d F\over dt}\Big|_0^\tau 
+{1 \over M^3} \int_0^\tau dt {d \sum_a \sigma_a \over dt}{d F\over dt}.
\end{align}
Omitting the surface term we find a metric of the form  
\begin{align}
g_{\mu\nu}^{SH}(\alpha)={1 \over 2M^3} 
\left(
{\p \sum_a \sigma_a \over \p \alpha^\mu} {\p F \over \p \alpha^\nu}
+{\p \sum_a \sigma_a \over \p \alpha^\nu} {\p F \over \p \alpha^\mu}
\right)
\end{align}
where $\alpha^\mu$ is the coordinate in which the functions $\sigma_a$ and $F$ transform as scalars.

We also examine the Hatano-Sasa entropy production:
\begin{align}
\langle \Sigma\rangle^{(k)}
&=
\left(-{1 \over M}\right)^{k+1}  \int_0^\tau dt \ln{M-F \over F} {d^{k+1} F\over dt^{k+1}} .
\end{align}
As stated in the previous section and we can check explicitly, the leading term $\langle \Sigma\rangle^{(0)}$ becomes the minus of the Shannon entropy difference.

The $k=1$ term in the Hatano-Sasa entropy production becomes
\begin{align}
\langle \Sigma\rangle^{(1)}
&=
{1 \over M^2} \int_0^\tau dt \ln{M-F \over F} {d^{2} F\over dt^{2}} 
\nonumber \\
&=
{1 \over M^2} \ln{M-F \over F} {d F\over dt}\Big|_0^\tau 
-{1 \over M^2} \int_0^\tau dt {d F\over dt}\frac{d}{dt}\left(\ln{M-F \over F}\right)  ,
\end{align}
from which we find a thermodynamic metric:
\begin{align}
g_{\mu\nu}^{HS}(\alpha)={1 \over MF(M-F)} {\p F \over \p \alpha_\mu} {\p F \over \p \alpha_\nu}.
\end{align}
The function $F$ should transform as a scalar in the coordinate $\alpha^\mu$.

\begin{figure}[h]
 \begin{minipage}[b]{0.50\linewidth}
  \centering
  \includegraphics[width=7.8cm, bb=0 0 1024 768]{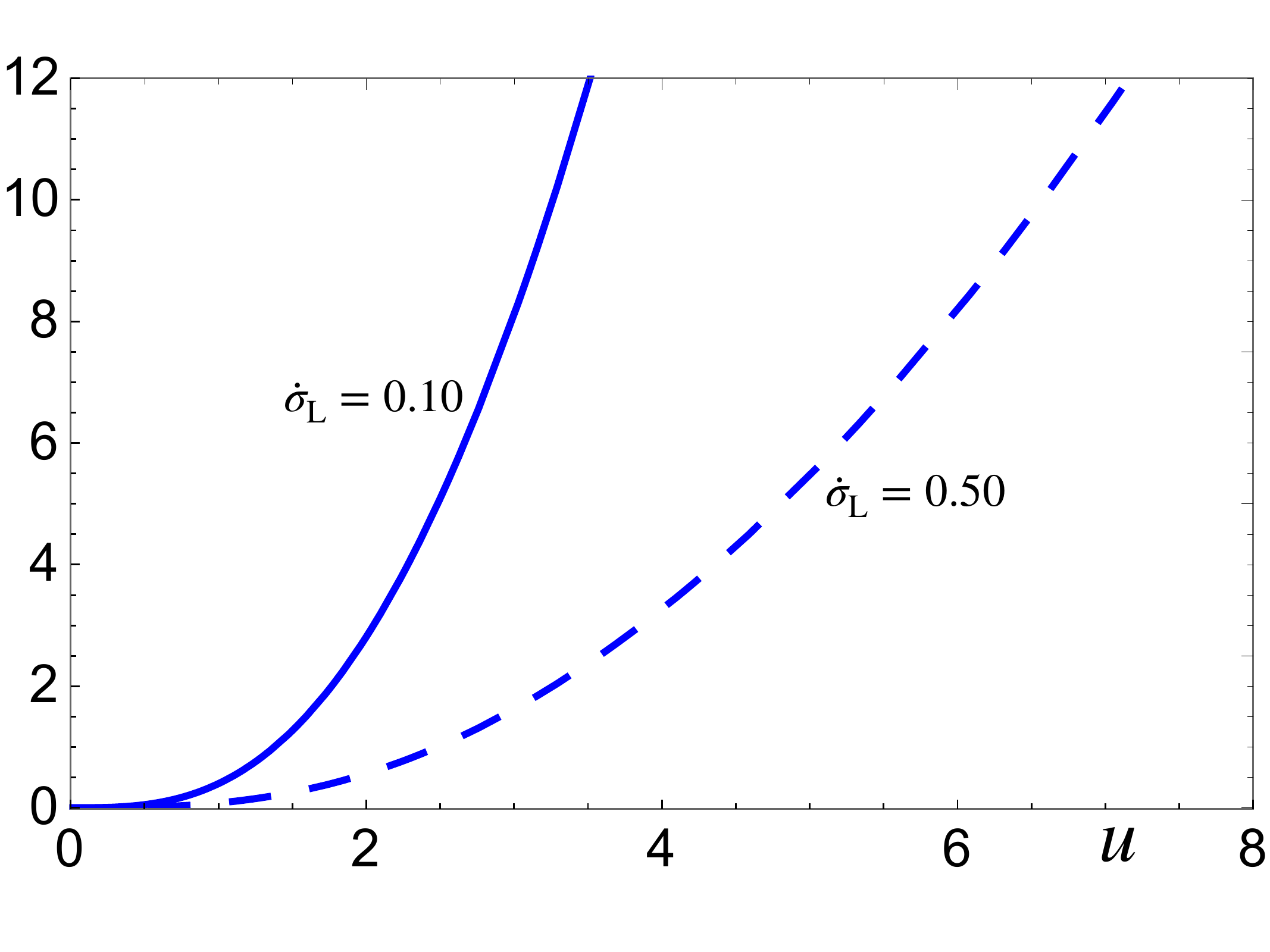}
  \subcaption{The housekeeping part of the entropy production.}
  \label{fig:housekeeping}
 \end{minipage}
 \begin{minipage}[b]{0.50\linewidth}
 \centering
  \includegraphics[width=7.8cm, bb=0 0 1024 768]{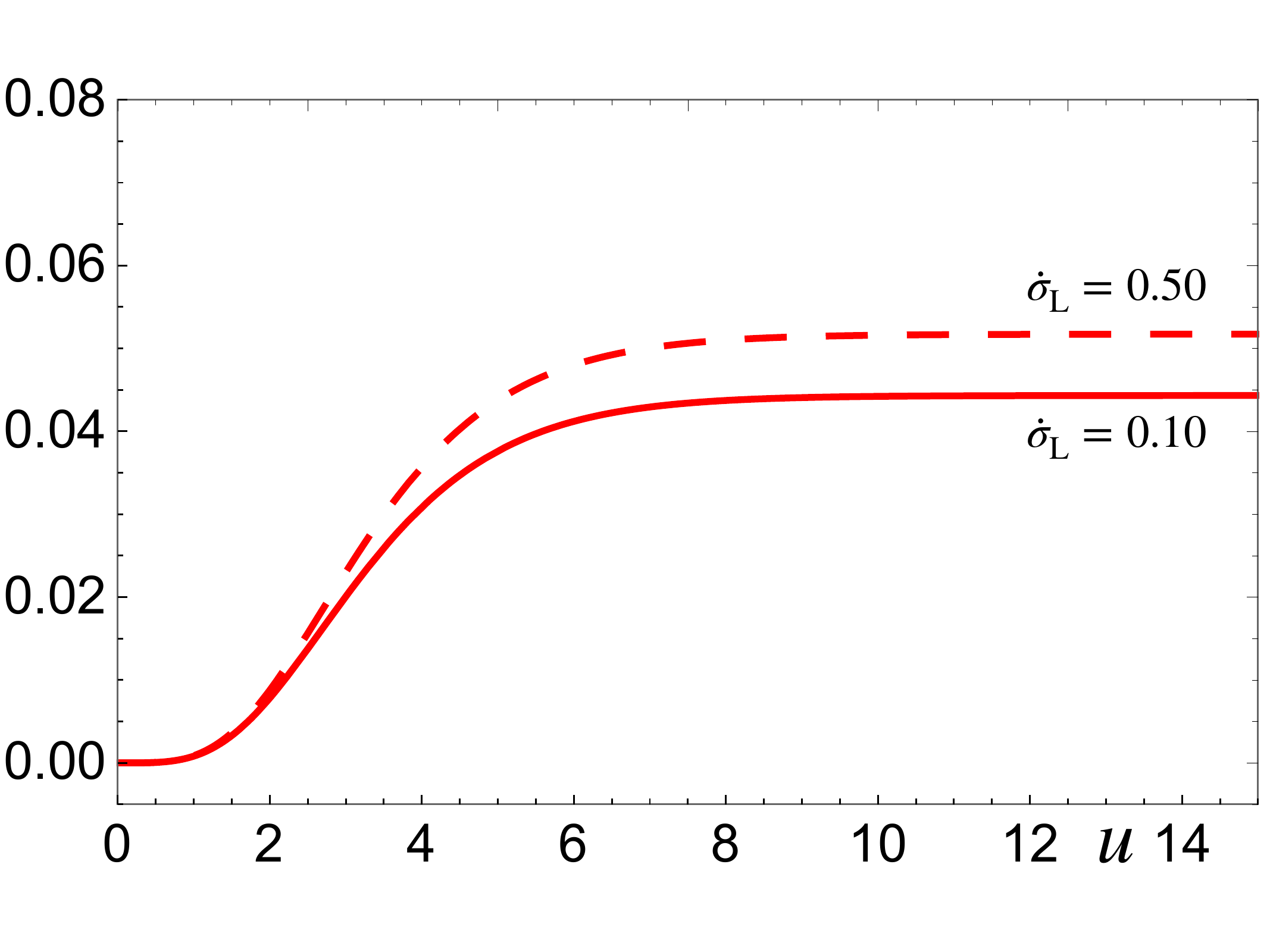}
  \subcaption{The difference between the housekeeping parts.}
  \label{fig:dfhk}
 \end{minipage}
 \\
 \begin{minipage}[b]{0.50\linewidth}
  \centering
  \includegraphics[width=8.5cm, bb=0 0 842 595]{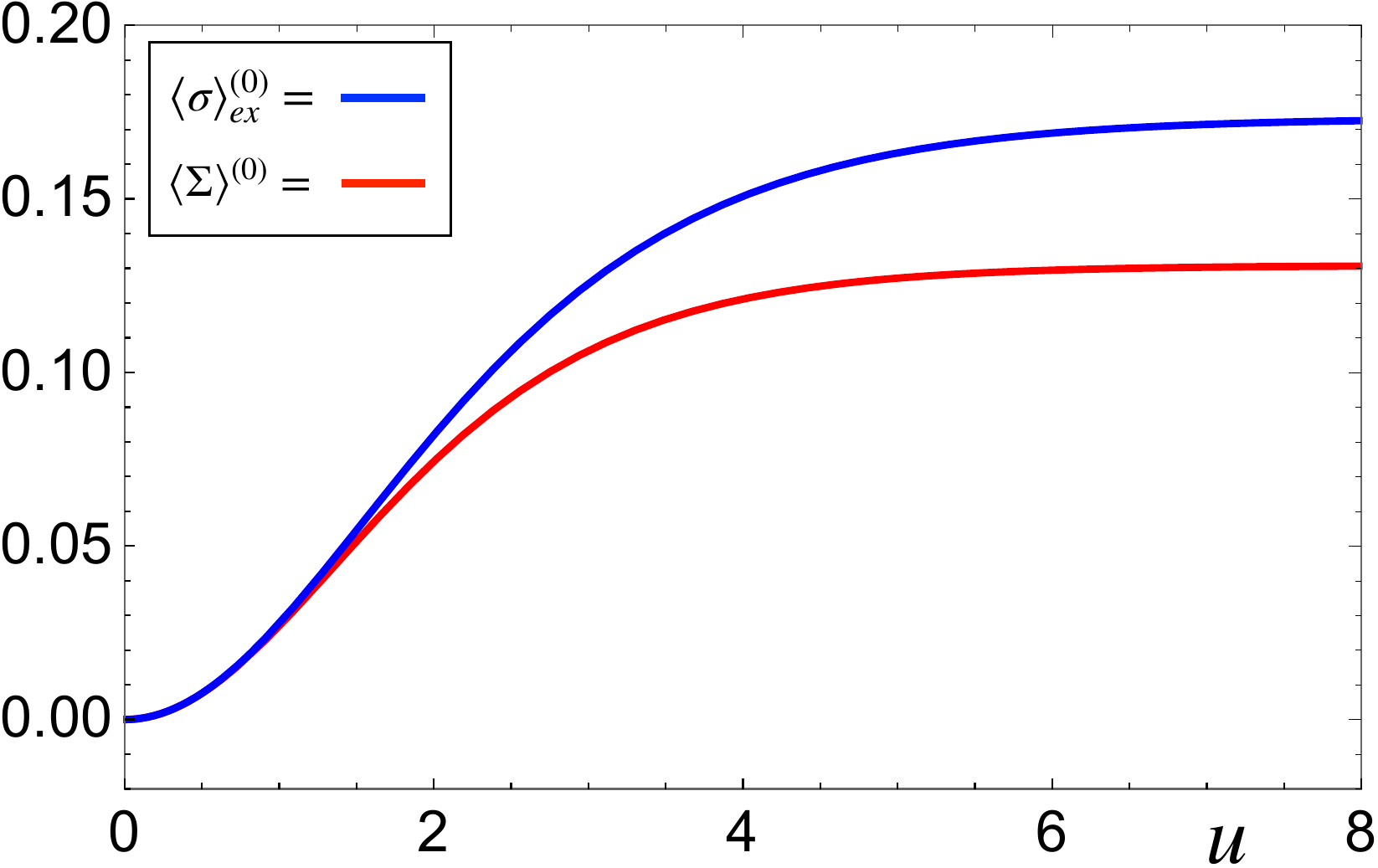}
  \subcaption{$k=0$ term of excess entropy production.}
  \label{fig:eep0term}
 \end{minipage}
 \begin{minipage}[b]{0.50\linewidth}
  \centering
  \includegraphics[width=8.5cm, bb=0 0 842 595]{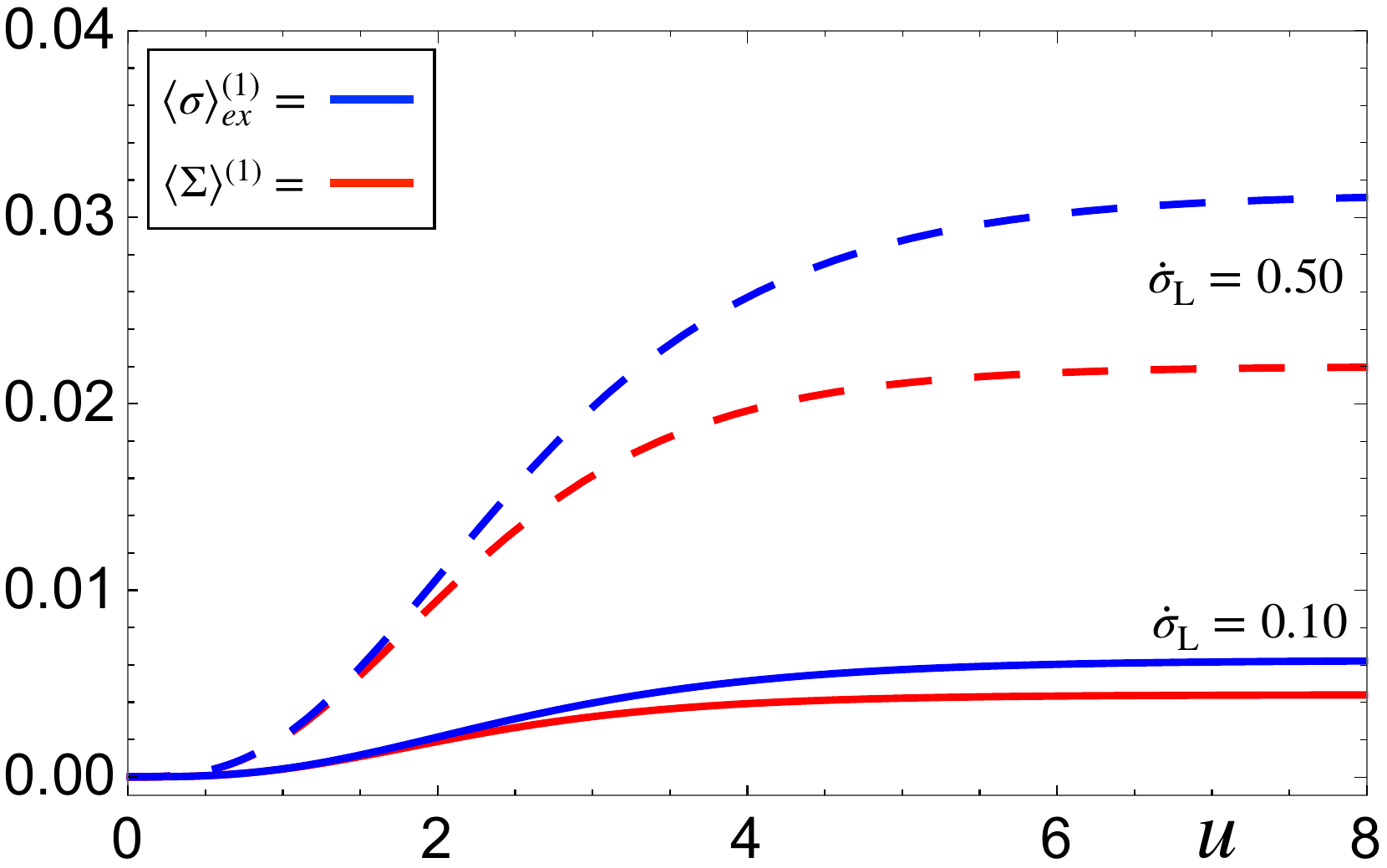}
  \subcaption{$k=1$ term of excess entropy production.}
  \label{fig:eep1term}
 \end{minipage}
 \caption{The behavior of the Sagawa-Hayakawa (blue lines) and the Hatano-Sasa (red lines) entropy production. }
 \label{fig:ep}
\end{figure}

As an illustration, we consider a protocol where the system couples to two reservoirs (left ($a=L$) and right ($a=R$)) and compare the behaviors of the two types of entropy production.
The dot is supposed to be initially in thermal equilibrium with $\sigma_L=\sigma_R=0$ corresponding to infinitely high temperature states.
Then we change $\sigma_L$ from $0$ to $u$, while $\sigma_R$ is left unchanged.
We assume the parameter $\sigma_L$ changes linearly in time $\sigma_L=b t$ where $b=\dot{\sigma}_L$ is a proportional constant. 
The explicit forms of the functions for the housekeeping parts of entropy production and the $k=0,1$ terms in the expansion are listed in the appendix and here we show the behaviors of these functions in Fig.\ref{fig:ep}.
In Fig.\ref{fig:housekeeping}, we see the housekeeping part of the Sagawa-Hakayawa entropy production diverges. 
As we see Fig.\ref{fig:dfhk}, there is a small difference between the housekeeping parts of the Sagawa-Hayakawa and Hatano-Sasa entropy production which we denote $\la \Delta\Sigma \ra_{hk}$ that approaches a constant at large $u$.
On the other hand, the excess parts of entropy production approach constants at large $u$ as we see in Fig.\ref{fig:eep0term} and Fig.\ref{fig:eep1term}. 
The expansion of $\langle \sigma\rangle^{(0)}_{ex}$ with respect to $u$ coincides with that of $\langle \Sigma\rangle^{(0)}$ up to $\mathcal{O}(u^2)$ and 
the behavior of $\langle \sigma\rangle^{(1)}_{ex}$ is the same as that of $\langle \Sigma\rangle^{(1)}_{ex}$ up to $\mathcal{O}(u^3)$. 
In Fig.\ref{fig:eep1term}, we see that the excess entropy production increases as the speed of the variation of the protocol $b$ increases.  
The $k=0, 1$ terms of the excess entropy production are positive, so the results satisfy a generalization of the second law of the SST for the transitions between NESSs \cite{RevModPhys.81.1665}. 

\section{Conclusion}

In this paper, we have investigated entropy production in finitely slow transitions between NESSs in stochastic master equations for Markov jump processes of finite states by using the improved adiabatic approximation which has a close relationship with the slow driving perturbation method.
Indeed, the former is obtained from the latter by omitting the off-diagonal transitions in the perturbation expansion in a particular gauge condition.
We focused on the Sagawa-Hayakawa and the Hatano-Sasa entropy production and
from the leading contributions of the expansion of the improved/slow driving approximations, which are valid in adiabatic/quasi-static processes, we confirmed the known results, namely the Berry-Sinitysn-Nemenman phase in the Sagawa-Hayakawa case and the Shannon entropy difference in the Hatano-Sasa case.
Furthermore, in the next leading order of the approximations, we obtained results in terms of thermodynamic metrics defined in parameter spaces.
However, we simply derived the thermodynamic metric without any applications. 
The physical implications should be uncovered in future works.
For the higher order of the approximations, we compared numerical results for the two types of the excess entropy production in a two-state system with a concrete protocol.
Although we studied the two-state system in which the improved adiabatic approximation is exactly the same as the slow driving perturbation, it is interesting to see any physical implication of the difference of the two methods. 
We are currently investigating the model which has more than two states and hope to report the results shortly.

\section*{Acknowledgement}
We are grateful to Kazutaka Takahashi for his useful discussions and comments.

\appendix
\section{Entropy production}
We show explicit expressions and asymptotic behaviors of entropy production in the protocol described in the text.

The housekeeping part of the Sagawa-Hayakawa entropy production becomes
\begin{align}
\la \sigma \ra_{hk}
&={1 \over 4b}
\int_0^u dx x \tanh\left({x \over 2}\right)
= \left \{
\begin{array}{l}
{1 \over 24b} u^3 - {1 \over 480b} u^5 + O(u^7) \quad (u \ll1)\\
\to \infty \quad (u \to \infty) ,
\end{array}
\right.
\label{sigma_dyn}
\end{align}
and the $k=0, 1$ terms of the entropy production become, respectively, 
\begin{align}
\la \sigma \ra_{ex}^{(0)}
&={1 \over 4}
\left[\ln 2-{u e^{-u}\over e^{-u}+1}-\ln (e^{-u}+1)
\right]
=\left \{
\begin{array}{l}
{1 \over 32} u^2 - {1\over 256} u^4 + O(u^6)\quad (u \ll1)\\
\to {1 \over 4} \ln 2 \quad (u \to \infty) ,
\end{array}
\right.
\label{sigma_ad}
\end{align}
and 
\begin{align}
\la \sigma \ra_{ex}^{(1)} 
&=
{b \over 8}
\left[
-{ue^{-u} \over (e^{-u}+1)^2}+{1 \over e^{-u}+1}-{1 \over 2}
\right]
=\left \{
\begin{array}{l}
{b \over 192} u^3 - {b \over 960} u^5 + O(u^7)\quad (u \ll1)\\
\to {b \over 16} \quad (u \to \infty) .
\end{array}
\right. 
\label{sigma_nad}
\end{align}

The housekeeping part of the Hatano-Sasa entropy production is
\begin{align}
\la \Sigma \ra_{hk}
&=
\la \sigma \ra_{hk}+ \la \Delta\Sigma \ra_{hk}
\nonumber \\
&=\left \{
\begin{array}{l}
{1 \over 24b} u^3 + {1\over 1024} u^4 + \left(-{1 \over 480b}+{b \over 5120}\right)u^5+O(u^6) \quad (u \ll1)\\
\to  \infty \quad (u \to \infty) ,
\end{array}
\right.
\label{HShk}
\end{align}
where
\begin{align}
\la \Delta\Sigma \ra_{hk}
&= 
\int_0^u dx 
{e^{-x}\left(2+b+(2-b)e^{-x}\right) \over 8(1+e^{-x})^3}
\left(x+2\ln{1+3  e^{-x} \over 3+ e^{-x}} \right)
\nonumber \\
&=\left \{
\begin{array}{l}
{1\over 1024} u^4 + {b \over 5120}u^5+O(u^6) \quad (u \ll1)\\
\to  {1 \over 16}\left\{20\ln2-8\ln3+3b(\ln3-1)\right\} \quad (u \to \infty) .
\end{array}
\right.
\label{HShk}
\end{align}
The $k=0, 1$ terms of the entropy production become, respectively,
\begin{align}
\la \Sigma \ra^{(0)}
&=
-\Delta H
\nonumber \\ 
&={2-F \over 2}\ln{2-F \over 2}+{F \over 2}\ln{F \over 2}+\ln 2
=\left \{
\begin{array}{l}
{1 \over 32} u^2 - {5\over 1024} u^4 + O(u^6) \quad (u \ll1)\\
\to  {3 \over 4} \ln 3 -\ln 2 \quad (u \to \infty) ,
\end{array}
\right.
\label{DSE}
\end{align}
where $F={1 \over e^{-u}+1}+{1\over 2}$, 
and
\begin{align}
\langle \Sigma\rangle^{(1)}
= \frac{be^u\left[(5+3\cosh{u})\ln\left(\frac{3+e^u}{1+3e^u}\right)+4\sinh{u}\right]}{8(1+e^u)^2}
=\left \{
\begin{array}{l}
{b \over 192} u^3 - {19 b \over 15360} u^5 + O(u^7) \quad (u \ll1)\\
\to  {b \over 16} (4 - 3\ln 3) \quad (u \to \infty) .
\end{array}
\right.
\end{align}
We have denoted the behaviors at small and large $u$ ($u \to \infty$ corresponds to zero temperature).
We see both of the housekeeping parts diverge and the excess parts approach constants at large $u$.


\begin{thebibliography}{99}

\bibitem{takahashi2020nonadiabatic}
K.~Takahashi, K.~Fujii, Y.~Hino and H.~Hayakawa,
``Nonadiabatic Control of Geometric Pumping,"
Phys.\ Rev.\ Lett.\  {\bf 124}, 150602 (2020)

\bibitem{Mandal_2016}
D.~Mandal and C.~Jarzynski,
``Analysis of slow transitions between nonequilibrium steady states,"
J.\ Stat.\ Mech.\  {\bf 2016}, 063204 (2016)

\bibitem{Cavina_2017}
V.~Cavina, A.~Mari, and V.~Giovannetti,
``Slow Dynamics and Thermodynamics of Open Quantum Systems,"
Phys.\ Rev.\ Lett.\  {\bf 119}, 050601 (2017)

\bibitem{PhysRevE.84.051110}
T.~Sagawa and H.~Hayakawa,
``Geometrical expression of excess entropy production,"
Phys.\ Rev.\ E\  {\bf 84}, 051110 (2011)

\bibitem{PhysRevLett.86.3463}
T.~Hatano and S.~Sasa,
``Steady-State Thermodynamics of Langevin Systems,"
Phys.\ Rev.\ Lett.\  {\bf 86}, 3463 (2001)



\bibitem{Seifert_2012}
U.~Seifert,
``Stochastic thermodynamics, fluctuation theorems and molecular machines,"
Rep.\ Pro.\ Phys.\  {\bf 75}, 126001 (2012)


\bibitem{Landauer_1978}
R.~Landauer,
``$dQ=T\mathrm{dS}$ far from equilibrium,"
Phys.\ Rev.\ A\  {\bf 18}, 255 (1978)

\bibitem{10.1143/PTPS.130.29}
Y.~Oono and M.~Paniconi,
``Steady State Thermodynamics,"
Prog.\ Theor.\ Phys.\ Suppl.\ {\bf 130}, 29 (1998)


\bibitem{Ruelle_2003}
D. P.~Ruelle,
``Extending the definition of entropy to nonequilibrium steady states,"
Proc.\ Natl.\ Acad.\ Sci.\ U.\ S.\ A.\ {\bf 100}, 3054 (2003)


\bibitem{KNST1}
T. S.~Komatsu, N.~Nakagawa, S.~Sasa and H.~Tasaki,
``Steady-State Thermodynamics for Heat Conduction: Microscopic Derivation,"
Phys.\ Rev.\ Lett.\  {\bf 100}, 230602 (2008)


\bibitem{Maes_2014}
C.~Maes and K.~Neto{\v c}n{\'y},
``A Nonequilibrium Extension of the Clausius Heat Theorem,"
J.\ Stat.\ Phys.\  {\bf 154}, 188 (2014)


\bibitem{dechant2022}
A.~Dechant, S.~Sasa and S.~Ito,
``Geometric decomposition of entropy production in out-of-equilibrium systems,"
arXiv:2109.12817



\bibitem{Crooks_1999}
G. E.~Crooks, 
``Entropy production fluctuation theorem and the nonequilibrium work relation for free energy differences,"
Phys.\ Rev.\ E\  {\bf 60}, 2721 (1999)




\bibitem{Speck_2005}
T.~Speck and U.~Seifert,
``Integral fluctuation theorem for the housekeeping heat,"
J.\ Phys.\ A: Math.\ Gen.\ {\bf 38}, L581 (2005)

\bibitem{Seifert_2005}
U.~Seifert,
``Entropy Production along a Stochastic Trajectory and an Integral Fluctuation Theorem,"
Phys.\ Rev.\ Lett.\  {\bf 95}, 040602 (2005)

\bibitem{Sasa_2006}
S.~Sasa and H.~Tasaki,
``Steady State Thermodynamics,"
J.\ Stat.\ Phys.\  {\bf 125}, 125 (2006)



\bibitem{KNST2}
T. S.~Komatsu, N.~Nakagawa, S.~Sasa and H.~Tasaki,
``Entropy and Nonlinear Nonequilibrium Thermodynamic Relation for Heat Conducting Steady States,"
J.\ Stat.\ Phys.\  {\bf 142}, 127 (2011)

\bibitem{Saito_2011}
K.~Saito and H.~Tasaki,
``Extended Clausius Relation and Entropy for Nonequilibrium Steady States in Heat Conducting Quantum Systems,"
J.\ Stat.\ Phys.\  {\bf 145}, 1275 (2011)

\bibitem{Pradhan_2011}
P.~Pradhan, R.~Ramsperger and U.~Seifert,
``Approximate thermodynamic structure for driven lattice gases in contact,"
Phys.\ Rev.\ E\  {\bf 84}, 041104 (2011)



\bibitem{PhysRevE.80.021137}
H.~Ge,
``Extended forms of the second law for general time-dependent stochastic processes,"
Phys.\ Rev.\ E\  {\bf 80}, 021137 (2009)

\bibitem{PhysRevE.81.051133}
H.~Ge and H.~Qian,
``Physical origins of entropy production, free energy dissipation, and their mathematical representations,"
Phys.\ Rev.\ E\  {\bf 81}, 051133 (2010)


\bibitem{Esposito_2010}
M.~Esposito and C.~Van den Broeck,
``Three Detailed Fluctuation Theorems,"
Phys.\ Rev.\ Lett.\  {\bf 104}, 090601 (2010)

%



\bibitem{RevModPhys.81.1665}
M.~Esposito, U.~Harbola, and S.~Mukamel,
``Nonequilibrium fluctuations, fluctuation theorems, and counting statistics in quantum systems,"
Rev.\ Mod.\ Phys.\  {\bf 81}, 1665 (2009)






\bibitem{CoverThomas}
T. M.~Cover and J. A.~Thomas,
``Elements of Information Theory,"
John Wiley and Sons, Inc., New York (2006)



\bibitem{Takahashi_2020}
K.~Takahashi, Y.~Hino, K.~Fujii and H.~Hayakawa,
``Full Counting Statistics and Fluctuation--Dissipation Relation for Periodically Driven Two-State Systems,"
J.\ Stat.\ Phys.\  {\bf 181}, 2206 (2020)



\bibitem{Berry1984}
M. V.~Berry,
``Quantal phase factors accompanying adiabatic changes,"
Proc.\ Roy.\ Soc.\ Lond.\ A {\bf 392}, 45 (1984)

\bibitem{PhysRevLett.58.1593}
Y.~Aharonov and J.~Anandan,
``Phase change during a cyclic quantum evolution,"
Phys.\ Rev.\ Lett.\  {\bf 58}, 1593 (1987)

\bibitem{PhysRevLett.60.2339}
J.~Samuel and R.~Bhandari,
``General Setting for Berry's Phase,"
Phys.\ Rev.\ Lett.\  {\bf 60}, 2339 (1988)

\bibitem{Sinitsyn_2007}
N. A.~Sinitsyn and I.~Nemenman,
``The Berry phase and the pump flux in stochastic chemical kinetics,"
Europhys.\ Lett.\ {\bf 77}, 58001 (2007)

\bibitem{PhysRevLett.99.220408}
N. A.~Sinitsyn and I.~Nemenman,
``Universal Geometric Theory of Mesoscopic Stochastic Pumps and Reversible Ratchets,"
Phys.\ Rev.\ Lett.\ {\bf 99}, 220408 (2007)

\bibitem{Nakajima_2021}
S.~Nakajima, and Y.~Utsumi
``Asymptotic expansion of the solution of the master equation and its application to the speed limit,"
Phys.\ Rev.\ E\  {\bf 104}, 054139 (2021)


\bibitem{articleOhkuboEggel}
J.~Ohkubo and T.~Eggel,
``Noncyclic and nonadiabatic geometric phase for counting statistics,"
J.\ Phys.\ A: Math.\ Theor.\ {\bf 43}, 425001 (2010)

%
%
%

\bibitem{PhysRevResearch.3.013187}
Y.~Hino and H.~Hayakawa,
``Geometrical formulation of adiabatic pumping as a heat engine,"
Phys.\ Rev.\ Research.\  {\bf 3}, 013187 (2021)


\bibitem{PhysRevB.67.085316}
D. A.~Bagrets and Y. V.~Nazarov,
``Full counting statistics of charge transfer in Coulomb blockade systems,"
Phys.\ Rev.\ B\  {\bf 67}, 085316 (2003)




%
%

\end{thebibliography}


\end{document}